\documentclass[11pt]{article}
\usepackage{epsfig}
\usepackage[usenames]{color}
\usepackage{amsmath}
\usepackage{amsfonts}
\usepackage{amssymb}
\renewcommand{\theequation}{\mbox{\arabic{section}.\arabic{equation}}}

\newtheorem{proposition}{Proposition}[section]
\newcommand{\bpr}{\begin{proposition}}
\newcommand{\epr}{\end{proposition}}

\newcounter{Roman}
\addtocounter{Roman}{1}

\newcommand{\beq}{\begin{equation}}
\newcommand{\eeq}{\end{equation}}
\newcommand{\bea}{\begin{eqnarray}}
\newcommand{\eea}{\end{eqnarray}}
\newcounter{saveeqn}

\newcommand{\D}{\displaystyle}

\newcommand{\ssc}{\scriptscriptstyle}

\newcommand{\hj}{\hat{\jmath}}

\newcommand{\bx}{{\bf x}}

\newcommand{\vev}[1]{\Big\langle #1 \Big\rangle}
\newcommand{\bpsi}{\bar{\psi}}

\newcommand{\C}{{\cal C}}

\input epsf
\begin{document}  

\begin{center}{\Large\bf  Chiral symmetry restoration at large chemical potential in 
strongly coupled $SU(N)$ gauge theories}\\[2cm] 
{E. T. Tomboulis\footnote{\sf e-mail: tomboulis@physics.ucla.edu}
}\\
{\em Department of Physics and Astronomy, UCLA, Los Angeles, 
CA 90095-1547}
\end{center}
\vspace{1cm}

\begin{center}{\Large\bf Abstract}\end{center} 
We show that at sufficiently large chemical potential $SU(N)$ lattice gauge theories in the strong coupling limit with staggered  fermions  are in a chirally symmetric phase. The proof employs a polymer cluster expansion which exploits the anisotropy between timelike and spacelike directions in the presence of a quark chemical potential $\mu$. The expansion is shown to converge in the infinite volume limit at any temperature for sufficiently large $\mu$. All expectations of chirally non-invariant local fermion operators vanish identically, or, equivalently, their correlations cluster exponentially, within the expansion. The expansion itself may serve as a computational tool at large $\mu$ and strong coupling. 

\vfill
\pagebreak

\section{Introduction} 
\setcounter{equation}{0}
\setcounter{Roman}{0}
Elucidating the rich structure of the QCD phase diagram, and more generally that of $SU(N_c)$ gauge theories, as a function of temperature and finite density continues to present a conceptual and technical challenge. A major obstacle in this endeavor has been the inability to perform lattice gauge theory ab initio simulations by standard Monte Carlo methods 
since the fermion determinant becomes complex in the presence of finite baryon chemical potential, the infamous ``sign problem". Significant progress has been made in recent years 
in overcoming some of these problems by a combination of numerical simulations, mostly for small chemical potential,  analytical techniques and investigation of model systems (see \cite{FH} \cite{dF} for review). 
Despite such progress, however, away from the region along the temperature axis 
the QCD phase diagram remains, for the most part, conjectural. In the absence of results established from first principles by simulations it is important to attempt to obtain any first principle results by other means.

In this paper we consider $SU(N_c)$ lattice gauge theory with $N_f$ flavors of massless staggered fermions in the strong coupling limit at temperature $T$ and quark chemical potential $\mu$.   
The theory has been investigated in the literature, in particular for the cases of $SU(2)$ and $SU(3)$ with  $N_f=1$, using a variety of approaches. 
Integrating out the gauge field at strong coupling leads to a representation of the partition function in terms of monomers, dimers and baryon loops \cite{DMW}, or monomers, dimers and polymers 
\cite{KM}, \cite{KIM}. In this representation the sign problem is partly evaded allowing simulations. In such simulations in the case of $N_c=3$ \cite{KM} a clear sign of a chiral symmetry restoring first order  transition was found at some critical $\mu$. Similarly, for $N_c=2$, restoration of chiral symmetry at large $\mu$ and/or $T$ was seen in  \cite{KIM}.  
More recently, the two-color ($N_c=2$), $N_f=1$ case was investigated in \cite{CJ} using the dimer-baryon loop representation with a new updating algorithm \cite{AC}. A second order transition to a chirally symmetric phase at some critical $\mu$ was seen in good agreement with  mean field predictions.  In the case of $N_c=3$ such improved simulations were carried out in \cite{FdF}. Another line of investigations follows  
a mean field approach. In  \cite{NFH}, \cite{N}, \cite{KMOO} an effective action was obtained by performing a $1/d$ expansion in the spatial directions and retaining only the leading terms, while leaving the timelike directions intact. Introducing auxiliary bosonization sources for condensates of interest, and finally integrating out the fermion and timelike gauge fields, treating condensates in a mean field approximation, leads to an effective action which allows a detailed study of the phase diagram in $T$, $\mu$ and quark mass $m$. This phase diagram exhibits a chiral phase at large $\mu$ and/or $T$.

Here we will obtain an actual general proof of the existence of a chirally symmetric phase for the  
$SU(N_c)$ gauge theory with $N_f$ flavors at strong coupling at chemical potential above a critical value. 
The basic approach adopted is as follows. 
Nonvanishing chemical potential and/or temperature introduce in the action an anisotropy between timelike and spacelike directions. This anisotropy grows with increasing $\mu$, and $T$, and 
provides the basis for setting up a systematic expansion in which the spacelike part of the action is expanded in a fermion hopping expansion in the measure provide by the timelike part. 

This type of expansion was first used in \cite{TY} to rigorously prove the restoration of chiral symmetry in lattice gauge theories at large $T$ (and $\mu=0$) for all values of the gauge coupling.  
Here we exploit the anisotropy to set up an expansion in the presence of nonvanishing $\mu$. 
For technical reasons \cite{F0}, however, the expansion has to be set up in a somewhat different way than in \cite{TY}. We only consider the strong coupling limit. 
(This restriction is actually related to the complex nature of fermion determinants.) The resulting expansion can be formulated as a polymer-type cluster expansion \cite{Cam}, \cite{BGJ}. 
The standard convergence criteria for polymer expansions can then be applied. One thus finds that 
the expansion converges in the infinite volume limit in spatial dimension $d\geq 1$ for chemical potential $\mu$ large enough.   
Within its region of convergence the expansion describes a chirally symmetric phase.   
 
In this paper we are only concerned with establishing the convergence of the expansion and thus the existence of a chirally symmetric phase at large chemical potential. In fact, the actual bounds we obtain (section 2.3 below) on the convergence region are rather non-optimal; they can be easily improved. We will not consider here the systematic evaluation of graphs in the expansion and/or resummation of classes of graphs, i.e.  application of the expansion as a computational tool. 
It is interesting to note in this connection that quantitative results obtained by retaining a leading approximation in our expansion can be expected to be similar to those in \cite{NFH}, \cite{N}, \cite{KMOO} where only the leading terms in a $1/d$ expansion in the spatial directions 
\cite{F01} are retained.

The paper is organized as follows. In section 2, after some preliminaries, the expansion is set up and its diagram structure laid out. We then formulate it as a polymer expansion of the logarithm of the partition function and of expectations of observables (section 2.2). Its convergence is examined in section 2.3. The preservation of chiral symmetry within the expansion follows then as a standard consequence of convergence of a cluster expansion (section 3). Some discussion and outlook for further work are given in section 4. The derivation of some explicit formulas used in the text is provided in the Appendix.

\section{The cluster expansion}
\setcounter{equation}{0}
\setcounter{Roman}{0}
\subsection{Preliminaries} 

The lattice $\Lambda$ is a $(d+1)$-dimensional periodic hypercubic lattice of size $L_s^d \times L$. It is convenient to always take the lattice lengths in 
Euclidean time ($L$) and space ($L_s$) to be even. $\Lambda$ may be taken anisotropic with timelike and spacelike lattice spacings $a_\tau$ and $a_s$, respectively. This is the standard way of allowing independent variations of couplings and the physical temperature  which is defined as $T=(La_\tau)^{-1}$. The spatial lattice obtained as a particular equal time slice of $\Lambda$ will be denoted by 
$\Lambda_s$. 

We use standard lattice gauge theory notations and conventions. Lattice site coordinates will be 
denoted by $x=(x^\lambda)= (x^0, {\bf x})$ with $\lambda=0,1,\ldots, d$, and 
${\bf x}=(x^j)$, $j=1,\ldots,d$. We also often write 
$x^0=\tau$. Lattice unit vectors in the space and time directions will be denoted by 
$\hat{\jmath}$ and $\hat{0}$, respectively. We generically denote lattice bonds by $b$, plaquettes by $p$, etc.  Bonds will be specified  more explicitly 
as $b=(x, \lambda)$, or $b_s=(x, j)=<x, x+\hat{\jmath}>$ if spacelike, and $b_\tau=(x, 0)=<x, x+\hat{0}> $ if timelike. Correspondingly, the gauge field variables $U_b$ defined on each $b\in \Lambda$ will, as usual, often be more explicitly specified by $U_j(x)$ and $U_0(x)$. 

The gauge fields $U_b\in G_c$ are elements of the gauge group (color) $G_c$ and taken to transform in the fundamental representation. Here $G_c=SU(N_c)$. Fermions are introduced by 
associating on each site $2\nu$ generators $\bpsi_\alpha(x)$, $\psi_\alpha(x)$, 
($\alpha=1,\ldots,\nu$), of a Grassmann algebra. The fermions are taken to transform as $N_f$ 
copies (flavors) of the fundamental representation of the gauge group.  
In explicit enumerations we use $a=1,\ldots, N_c$, $i=1,\ldots,N_f$ for color and flavors indices. We assume periodic boundary conditions for the gauge field and antiperiodic boundary conditions for the fermions. 

We consider fermions in the presence of chemical potential $\mu/ a_\tau$. 
The lattice action for massless fermions is then given by: 
\bea
S_F & = & 
\sum_{b_s=(x,j)}{1\over 2}\left({a_\tau\over a_s}\right)  \,\left[ \bar{\psi}(x) \gamma_j(x) U_j(x) \psi(x+\hat{\jmath}) - \bar{\psi}(x+\hat{\jmath}) \gamma_j(x) U^\dagger_j(x) \psi(x) \right]  \nonumber \\
  & + & \!  \sum_{b_\tau=(x,0)} {1\over 2} \,\left[ \bar{\psi}(x) \gamma_0(x) e^{\mu } U_0(x) \psi(x+\hat{0}) - \bar{\psi}(x+\hat{0}) \gamma_0(x) e^{-\mu } U^\dagger_0(x) \psi(x) \right]  .  \label{act1}
\eea
The matrices $\gamma[b]=\gamma_\lambda(x)$ defined on each bond $b=(x,\lambda)$ 
satisfy $\prod_{b\in \partial p}\gamma[b]=1$ for each plaquette $p$. 
For staggered fermions 
\beq
\gamma_\lambda(x) = (-1)^{\sum_{\nu< \lambda} x^\nu}  \, ,   \qquad \gamma_0(x) = 1 \, .\label{gamma1} 
\eeq
For staggered fermions then one has $\nu=N_cN_f$, and thus $\alpha=(a,i)$. Recall that $N_f$ staggered flavors correspond to $4N_f$ continuum flavors. 
For naive fermions $\gamma_\lambda(x) = \gamma_\lambda$ are elements of the 
Euclidean Dirac-Clifford algebra satisfying $\{\gamma^\lambda, \gamma^\kappa\} = 2 \delta^{\lambda\kappa}{\bf 1}$ and are hermitian matrices. Naive fermions are of course unitarily equivalent to $2^{(d+1)/2}$ copies of staggered fermions. We only consider staggered fermions in the following even though we often state formulas for general $\gamma[b]$'s. 

With $N_f$ flavors of staggered fermions the fermion action in (\ref{act1}) is  invariant under a $U(N_f)\times U(N_f)$ global symmetry corresponding to independent rotations of fermions on even and odd sublattices \cite{F1}. Explicitly, for any element $(u,v)\in U(N_f)\times U(N_f)$, 
they are given by $\psi(x) \to u\psi(x)$ and $\bpsi(x) \to \bpsi(x)v^\dagger$ for even sites 
and $\psi(x) \to v \psi(x)$ and $\bpsi(x) \to \bpsi(x)u^\dagger$ for odd sites. 
This symmetry is referred to as chiral symmetry. 
  
It is the presence of this chiral symmetry that dictates the use of staggered (naive) fermions 
here. The cluster expansion with staggered fermions employed in the following can be set up equally well for Wilson fermions. The latter, however, explicitly break the chiral symmetry (to be recovered only in the continuum). Convergence of the expansion with Wilson fermions at finite lattice spacing then does not allow one to directly conclude anything away from the continuum limit. 
Domain wall fermions would certainly be an interesting alternative, but, due to the more complicated, non-local nature of their action, will not be considered in this paper \cite{Fim}. 

We work in the strong coupling limit, i..e. at inverse gauge coupling $\beta=0$, so a 
gauge field plaquette action term is absent. The full measure is then given by 
\beq 
d\mu_\Lambda = Z_\Lambda^{-1} \,\prod_{b\in \Lambda} dU_b \prod_{x\in \Lambda} d\bpsi(x) d\psi(x) \exp(-S_F) \, , \label{meas1}
\eeq
where, as usual, $dU_b$ denotes normalized Haar measure on the group $G_c$, and 
$d\bpsi(x) d\psi(x)\equiv \prod_\alpha d\bpsi_\alpha(x) d\psi_\alpha(x)$ is the standard measure on a Grassmann algebra. The partition function $Z_\Lambda$ is defined by $\int d\mu_\Lambda=1$. 
The measure (\ref{meas1}) is invariant under the global chiral symmetry (as well as, of course, under the local $G_c$ gauge symmetry). 

Expectations of general fermionic observables ${\cal O}$ are then given by 
\beq 
\vev{{\cal O}} = \int d\mu_\Lambda \; {\cal O}  \,. \label{exp1}
\eeq
We are interested, in particular, in the possible formation of condensates of products of local fermion operators such as the chiral order parameter $\bpsi(x)\psi(x)$ and various diquark or multi-quark operators. 
Introducing the notation $\chi(x) \equiv \left(\begin{array}{c} \psi(x) \\ \bpsi(x)^T 
\end{array} \right)$, such operators are of the general form 
\beq 
{\cal O} = \prod_{x\in O} \chi(x)^T \Gamma_x \chi(x)  \equiv \prod_{x\in O} {\cal O}(x)  
\,, \label{exp2}  \eeq 
where $O$ (the support of ${\cal O}$) is a finite set of sites, and $\Gamma_x$ is some (unitary) matrix. 

\subsection{The expansion} 

The presence of a nonvanishing chemical potential in (\ref{act1}) introduces an anisotropy between the spacelike and timelike directions which can be exploited, in a manner analogous to that of the case of large $T$ \cite{TY}, to set up a convergent expansion for large $\mu$. For this 
purpose we rewrite (\ref{act1}) more concisely in the form
\beq 
S_F = \sum_{x,y} \bpsi(x)  {\bf M}^{(s)}_{xy}(U) \psi(y)  + 
 \sum_{x,y} \bpsi(x) {\bf M}^{(t)}_{xy}(U)  \psi(y) \, ,\label{act2}  
\eeq 
where the matrices ${\bf M}^{(s)}$ and ${\bf M}^{(t)}$ have nonvanishing elements 
only between nearest-neighbor sites, i.e., on bonds, given by
\beq
{\bf M}^{(s)}_{x (x+\hat{\jmath})}(U) = {1\over 2} (a_\tau/ a_s ) 
  \gamma_j(x) U_j(x)  \qquad \mbox{and} \qquad   
  {\bf M}^{(s)}_{(x+\hat{\jmath})x}(U) = -{1\over 2} 
  (a_\tau/ a_s)  \gamma_j (x)U^\dagger_j(x)  \label{Ms}
  \eeq 
for spacelike neighbors, and 
\beq 
{\bf M}^{(t)}_{x, x+\hat{0}}(U) = {1\over 2} \gamma_0 e^{\mu } U_0(x)
\qquad \mbox{and} \qquad {\bf M}^{(t)}_{x+\hat{0},x}(U) = -{1\over 2} \gamma_0 e^{-\mu } 
U^\dagger_0(x) \label{Mt} 
\eeq
for timelike neighbors.

To set up our expansion we rewrite the measure (\ref{meas1}) as follows.   
For each spacelike bond $b_s=<x,x+\hj>$ and each $\alpha$ let 

\beq
f_{b_s}^{1,\alpha} \equiv 
\bpsi_\alpha(x) ({\bf M}^{(s)}_{x (x+\jmath)}\psi(x+\jmath) )_\alpha \;, 
\qquad 
f_{b_s}^{2, \alpha} \equiv (\bpsi(x+\jmath) {\bf M}^{(s)}_{(x+\jmath)x})_\alpha \psi_\alpha(x)  
\;. \label{fdef}    
\eeq
Note that $(f_{b_s}^{l,\ \alpha})^2=0$. One then has 
\beq 
\exp \left( 
\bpsi(x) {\bf M}^{(s)}_{x(x+\hj)}(U) \psi(x+\hj) + 
 \bpsi(x+\jmath) {\bf M}^{(s)}_{(x+\hj)x}(U) \psi(x)\right) 
= \
\prod_{l=1}^2 \prod_{\alpha=1}^{\nu} (1 +  f_{b_s}^{l, \alpha})  \; . \label{frep} 
\eeq

For each site $\bx$ in a fixed time slice $\Lambda_s$ define the measure 
\beq 
d\mu_{\bx}  =  {1\over z } \prod_{\tau=1}^L dU_0(\tau, \bx) 
d\bpsi(\tau,{\bx}) d\psi(\tau,{\bx})
\exp\left( \sum_{\tau, \tau^\prime} \bpsi(\tau,\bx) {\bf M}^{(t)}_{(\tau,\bx) (\tau^\prime,\bx)}(U_0)  \psi(\tau^\prime,\bx)  \right)  
\label{dmux}
\eeq
Note that 
\bea
& &  
 \sum_{\tau, \tau^\prime} \bpsi(\tau,\bx) {\bf M}^{(t)}_{(\tau,\bx) (\tau^\prime,\bx)}(U_0)  \psi(\tau^\prime,\bx)     \nonumber \\
& = &  
\sum_{\tau=1}^L[ \bpsi(\tau,\bx) \gamma_0 e^{\mu }U_0(\tau,\bx)\psi(\tau+1,\bx) - 
\bpsi(\tau+1,\bx) \gamma_0 e^{-\mu}U^\dagger_0(\tau,\bx)\psi(\tau,\bx)  \label{dmuxact1}
\eea
with $\psi(L+1,\bx) \equiv - \psi(1,\bx)$, $\bpsi(L+1,\bx) \equiv - \bpsi(1,\bx)$ due to the antiperiodic fermion boundary conditions. 
The factor $z$, defined by $\int d\mu_{\bx}=1$, is then 
a partition function along a  1-dimensional timelike fermion chain given by 
\beq
z = \int \prod_{\tau=1}^L dU_0(\tau, \bx) \, {\rm Det}{\bf M}^{(t)}_\bx(U_0) 
\label{z}
\eeq
with ${\bf M}^{(t)}_{\bx}(U_0) \equiv \left( {\bf M}^{(t)}_{(\tau,\bx) (\tau^\prime,\bx)}(U_0)\right)$  denoting the restriction of ${\bf M}^{(t)}_{xy}(U_0)$ 
to the submatrix of timelike bonds at fixed $\bx$. 

Using (\ref{frep}) and (\ref{dmux}) the full measure (\ref{meas1}) can now be expressed in the form
\beq 
d\mu_\Lambda = \tilde{Z}_\Lambda^{-1} \prod_{\bx \in \Lambda_s} d\mu_{\bx} 
\prod_{b_s\in \Lambda} dU_{b_s} 
 \prod_{b_s\in \Lambda}
 \prod_{l=1}^2  \prod_{\alpha=1}^{\nu} (1 +  f_{b_s}^{l, \alpha}) \, \label{meas2}
\eeq
with 
\beq 
\tilde{Z}_\Lambda \equiv Z_\Lambda \slash z^{|\Lambda_s|}  \,. \label{tildeZ} 
\eeq
The timelike part of the measure (\ref{meas2}) factorizes in a product with each factor representing the fermionic degrees of freedom coupled in a 1-dimensional timelike chain at fixed spatial coordinates $\bx$.  It is very convenient then to adopt the gauge, frequently referred to as Polyakov gauge, where the bond variables $U_0(\tau, \bx)$ are independent of $\tau$ and diagonal: 
\bea
U_0(\tau, \bx) & = & {\rm diag} (e^{i\theta_1(\bx)/L}, e^{i\theta_2(\bx)/L}, \cdots, 
e^{i\theta_{N_c}(\bx)/L} ) 
\; . \label{Polgauge1}\\
& = & \exp (i\Theta(\bx)/L)   \label{Polgauge2}
\eea
with 
\beq 
\sum_{a=1}^{N_c} \theta_a(\bx) = 0   \qquad \mbox{and } \qquad  \Theta(\bx) \equiv {\rm diag}( 
\theta_1(\bx), \theta_2(\bx), \cdots, \theta_{N_c}(\bx)) \label{Polgauge3} \,.
\eeq
This is always possible by gauge transformations setting the $U_0$'s at fixed spatial coordinates $\bx$ equal and rotating into the Cartan subgroup \cite{F2a}, \cite{F2b}.  After a time Fourier transform the 
timelike action  (\ref{dmuxact1})  becomes 
\beq 
 \sum_{\tau, \tau^\prime} \bpsi(\tau,\bx) {\bf M}^{(t)}_{(\tau,\bx) (\tau^\prime,\bx)}(U_0)  \psi(\tau^\prime,\bx)  = 
 \sum_{k_m \in {\sf BZ}} \bpsi(k_m,\bx) i \gamma_0 \sin \Big( k_m + 
 \Theta(\bx)/L -i\mu  \Big) \psi(k_m, \bx) \,, \label{dmuxact2}
 \eeq
and, hence, the  timelike propagator (covariance) in the background of the gauge field (\ref{Polgauge2}) is 
given by 
\beq 
C_{\tau, \tau^\prime}( \Theta(\bx)) \equiv C(\tau - \tau^\prime, \Theta(\bx)) = {1\over L} \sum_{k_m\in {\sf BZ}}e^{\D i k_m (\tau - \tau^\prime)}  C(k_m, \Theta(\bx) ) 
\label{prop1}
 \eeq 
 with  
 \beq
 C(k, \Theta(\bx) ) =  \left[ i\gamma_0  \sin \Big( k +  \Theta(\bx)/L -i\mu  \Big)\right]^{-1} \, . 
 \label{prop2}
 \eeq
In (\ref{dmuxact2}) and (\ref{prop1})  the summation is over momenta in the Brillouin zone ({\sf BZ}): 
 \beq 
 k_m = (2m-1) \pi / L \;, \quad  -L/2 + 1 \leq  m  \leq L/2  \qquad \Rightarrow  \quad 
 -\pi + \pi/L \leq k_m \leq \pi - \pi/L    \label{BZ} 
 \eeq 
(integer $m$). Evaluation of (\ref{prop1}) (Appendix A) gives 
\bea
C(\tau - \tau^\prime, \Theta(\bx))_{ai,bj} & = & \delta_{ab}\delta_{ij} [1 - (-1)^{\D (\tau-\tau^\prime)}] \, 
{  e^{\D -i\theta_a(\bx) (\tau-\tau^\prime)/L}  \over 
1 +e^{\D -i\theta_a(\bx) }  e^{ \D - \mu L}  } \, e^{\D - \mu (\tau - \tau^\prime)} \nonumber \\
& &\nonumber \\
& &  \quad \qquad \qquad \qquad\qquad  \mbox{for} \quad
  (\tau-\tau^\prime) > 0, \quad \mu >0 
\label{prop3a}
\eea
and 
\bea
 C(\tau - \tau^\prime, \Theta(\bx))_{ai,bj} 
& = & \! -\delta_{ab}\delta_{ij} [1 - (-1)^{\D |\tau-\tau^\prime|}] \, 
{e^{\D -i\theta_a(\bx)[1 -  |\tau-\tau^\prime|/L]} 
 \over  1 +e^{\D -i\theta_a(\bx) } e^{\D  - \mu L}  } \, 
 e^{ \D - \mu[L -  |\tau - \tau^\prime|]} \nonumber \\
 & & \nonumber \\
 & & \qquad  \qquad \qquad \qquad \qquad \qquad 
\mbox{for} \quad
  (\tau-\tau^\prime) < 0, \quad \mu >0 \; .
\label{prop3b}
\eea
For $\mu < 0$ (i.e., nonvanishing antiquark chemical potential)  replace $\mu, \theta_a(\bx)$ by $|\mu|, -\theta_a(\bx)$,  and reverse the sign condition on $(\tau - \tau^\prime)$ in (\ref{prop3a})-(\ref{prop3b}).

Note that $C(\tau, \Theta(\bx))$  vanishes for even $\tau$. This is a consequence of the chiral invariance of the action. The other salient property of $C(\tau, \Theta(\bx))$ is its 
exponential decay for nonvanishing $\mu$. This decay is in fact the crucial property needed for the convergence of the expansion below. It is worth pointing out that it can be extracted rather 
simply from (\ref{prop1}), without its actual evaluation, as follows. 
Consider the quantity 
\beq 
\tilde{C}(\tau - \tau^\prime, \Theta(\bx)) = 
\int_{-\pi}^\pi  {dk\over 2\pi} \; e^{\D i k (\tau - \tau^\prime)} 
C (k, \Theta(\bx))    \label{tprop}
 \eeq 
obtained by replacing the Brillouin zone in (\ref{prop1}) with the interval $[-\pi,\pi]$, i.e., 
its infinite $L$ (zero temperature) limit. Now, $C(k, \Theta(\bx))$ is analytic in $k$ for 
$|{\rm Im}\,  k|< |\mu|$ and bounded uniformly in ${\rm Re\,} k \in [-\pi,\pi]$. The periodicity of the integrand then allows one to ``lift" the contour of integration in (\ref{tprop}) (a standard trick) from the real axis to, say, ${\rm Im\,} k ={1\over 2} |\mu|$ or $-{1\over 2} |\mu|$,  which immediately establishes the exponential fall-off of (\ref{tprop}) and the bound 
\beq 
|\tilde{C}(\tau, \Theta(\bx))| \leq (\sinh (|\mu|/2))^{-1} \exp({\D -{|\mu|\over 2} |\tau|}) \label{Opropbound} 
\eeq
in all cases. The exponential decay of $C(\tau, \Theta(\bx))$ then follows from the relation (see (\ref{Aproprel})):  
\beq 
C( \tau, \Theta(\bx)) =\sum_{n=0}^\infty  (-1)^n \tilde{C}( \tau +nL, \Theta(\bx))  \,. 
\label{proprel} 
\eeq 
Explicit evaluation, i.e., eqs. (\ref{prop3a}), (\ref{prop3b}), results in a better bound, but, if all 
one is interested in is demonstrating convergence at sufficiently large $\mu$, the bound 
(\ref{Opropbound}) is sufficient. We also note (cf. Appendix A) that evaluation of (\ref{tprop}) and use of (\ref{proprel}) allows an alternative, simpler derivation of (\ref{prop3a})-(\ref{prop3b}) than the direct evaluation of (\ref{prop1}). 

We also need the quantity ${\rm Det}{\bf M}^{(t)}_\bx(U_0)$ resulting from integration of the fermions along the timelike chain at fixed spatial coordinates $\bx$ in (\ref{dmuxact1}). 
Its straightforward evaluation in the gauge (\ref{Polgauge2}) (Appendix A) gives: 
\bea 
{\rm Det}{\bf M}^{(t)}_\bx(U_0) & = & {\rm Det} C^{-1}(\Theta(\bx)) 
  \nonumber \\
 & = &  2^{-\nu L} e^{ \nu\mu L}  \left( \prod_{a=1}^{N_c} \, \Big[ 1 + e^{\D -i\theta_a}
 e^{\D - \mu L}\Big]^2 \right)^{N_f} \, . \label{Det} 
\eea
Note that in the case of $N_c=2$ one has $\theta_1=-\theta_2\equiv \theta$ and (\ref{Det}) becomes 
\beq 
{\rm Det} C^{-1}(\Theta(\bx)) =   2^{-\nu (L-1)} [ \cos\theta + \cosh L \mu]^{2N_f}
\, , \label{Detsu2} 
\eeq
which is indeed real and non-negative.

Our expansion is generated by expanding the products over spacelike bonds in  (\ref{meas2}) around the measure provided by the timelike part in (\ref{meas2}). In other words, we will perform a hopping expansion in the spacelike directions with the spacelike hops connected in the timelike directions by the propagators (\ref{prop3a})-(\ref{prop3b}). In the following we set $a_\tau/a_s=1$ for convenience.

Expanding first the products over the spacelike bonds gives
\beq 
\tilde{Z}_\Lambda = \sum_B \int \prod_{\bx \in \Lambda_s} d\mu_{\bx} 
\prod_{b_s \in B} dU_{b_s} \prod_{b_s, l, \alpha \in B} f_{b_s}^{l,\alpha} \label{Zexp1}
\eeq
The sum is over all subsets $B$ of the set of $2\nu$ copies of the set of space-like bonds in 
$\Lambda$. 
Each such $B$ may be decomposed into a number of connected components where connectivity is defined as follows: two elements of a set $B$ are connected if
they may be joined by a sequence of time-like bonds. 
Corresponding to this decomposition of each set $B$, and after carrying out fermion and gauge field integrations, each term in (\ref{Zexp1}) decomposes into 
a product or a sum of products \cite{F3} of factors, each factor being the value of a connected diagram. 

\begin{figure}[ht]
\begin{center}
\includegraphics[width=0.7\textwidth]{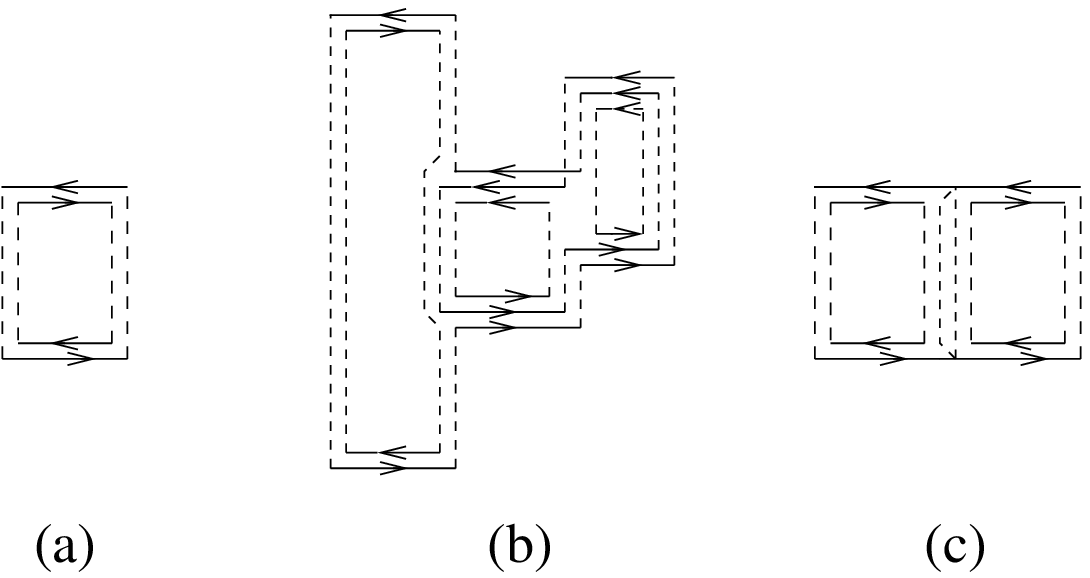}
\end{center}
\caption{Examples (a) - (c) of connected diagrams built of spacelike bonds (solid lines) 
connected by timelike propagators (broken lines). An equal mod $N_c$ number of $U$ and $U^\dagger$ (indicated by arrows) must occur on every spacelike bond ($N_c=3$ in (b)). Note that diagrams such as the diagram in  (c) cannot be factored in the product of two diagrams (a) since propagators along a common set of timelike bonds are ``welded" together by the integration over the background gauge field residing on them (cf. (\ref{polactiv2})). \label{cpcrF1}}
\end{figure}

A connected diagram $\gamma$ consists of a number of  space-like bonds connected by $1$-dimensional chains of 
time-like bonds. It is important that 
the number of $f^{l,\alpha}_b$ residing on each spacelike bond $b_s$ in a given diagram must result in an equal mod $N_c$ number of $U_{b_s}$ and $U^\dagger_{b_s}$'s on $b_s$ in 
order to obtain a nonvanishing result upon performing the $U_{b_s}$ integrations. This is, of course, a characteristic feature  of any fermion hopping expansion at strong coupling (absence of plaquette term). Examples are shown in Fig. \ref{cpcrF1}. 

\begin{figure}[ht]
\begin{center}
\includegraphics[width=0.35\textwidth]{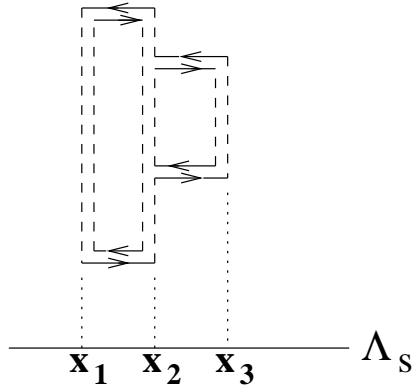}
\end{center}
\caption{The set $\{\bx \in \gamma\}$, in this example $\{\bx_1, \bx_2,\bx_3\}$, belonging to a given connected diagram $\gamma$ (cf. text).    \label{cpcrF2}}
\end{figure}

Each such connected diagram  $\gamma$ 
will be termed  a polymer. The value of the diagram $\zeta(\gamma)$ is the activity of the polymer $\gamma$: 
\bea 
\zeta(\gamma)&  = & \int \prod_{{\bf x}\in \gamma} d\mu_{\bf x} 
\prod_{b_s \in \gamma} dU_{b_s} \;  \prod_{b_s, l, \alpha\, \in \,\gamma } f^{l,\alpha}_{b_s} 
\label{polactiv1} \\
& = &( \prod_{{\bf x}\in \gamma} z^{-1}) \int \, \prod_{{\bf x}\in \gamma} dU_0(\Theta({\bx})) \, \prod_{{\bf x}\in \gamma}
{\rm Det} C^{-1}(\Theta(\bx)) \prod_{{\bf x}\in \gamma} \prod_{(\tau_i \tau_j)}  C_{\tau_i\tau_j} (\Theta(\bx))  \; I_\gamma 
\, .  \label{polactiv2}  
 \eea
Here $\{\bx \in \gamma\}$ denotes the set of spatial sites obtained by projecting $\gamma$ 
onto a spacelike slice $\Lambda_s$ (cf. Fig. \ref{cpcrF2}). 
In (\ref{polactiv2}) 
$(\tau_i\tau_j)$ denotes pairs of points along a timelike bond chain at $\bx$ connected by propagators, and $I_\gamma$ is a $N_c, N_f$-dependent constant resulting from the  fermion and $U_{b_s}$ integrations in (\ref{polactiv1}).  
The expansion (\ref{Zexp1}) then becomes: 
\beq
\tilde{Z}_\Lambda = 1 + \sum_{k=1}^\infty  \; {1\over k!} \sum _{(\gamma_1, \ldots, \gamma_k) \in {\cal D}_k } \; \prod_{i=1}^k \zeta(\gamma_i)  \; ,\label{tildeZ3}
\eeq
where ${\cal D}_k$ denotes the set of all sets $(\gamma_1, \gamma_2, \ldots, \gamma_k)$ of $k$ disjoint polymers. In (\ref{tildeZ3}) we conveniently sum over all ordered sequences of polymers and, correspondingly, divide by $k!$.

\begin{figure}[ht]
\begin{center}
\vspace{0.4cm}  
\includegraphics[width=0.6\textwidth]{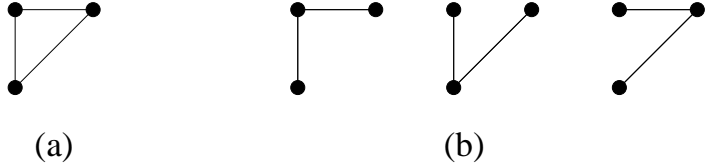}
\end{center}
\caption{(a) The graph $G(X)$ of a  cluster $X=\{\gamma_1,\gamma_2,\gamma_3\}$  consisting of three mutually intersecting polymers. (b) The set of all connected (proper) subgraphs  on $X$. \label{cpcrF3}} 
\end{figure}

The log of $\tilde{Z}_\Lambda$ is now obtained through an application of the moment-cumulant formalism of statistical mechanics. Let $X = (\gamma_1, \ldots, \gamma_k)$  
be a set of $k$ polymers (not necessary distinct); we will refer to such as set as a $k$-polymer ($k$-cluster). To each such $X$ associate a $k$-vertex graph G \cite{F4} as follows. Each vertex represents one $\gamma_i \in X$, and two vertices representing $\gamma_i$ and $\gamma_j$ are connected by a line if they intersect (when identified with subsets of the lattice $\Lambda$ as described above). An example is shown in Fig. \ref{cpcrF3}(a). 
 A  polymer occurring with multiplicity $m$, ($1\leq m\leq k$), in $X$ contributes $m$ vertices pairwise connected with lines (intersects itself). 
Thus the set ${\cal D}_k$ of all disjoint $k$-polymers in (\ref{tildeZ3}) 
is the set of all $X$ whose associated graphs consist of $k$ disconnected vertices (no lines). 
The set of all connected $k$-polymers, i.e. those $X$ such that no $\gamma_i\in X$ does not intersect at least one other $\gamma_j\in X$, will be denoted by $\C_k$. Thus the set $\C_k$ consists of all those clusters $X$ whose associated graphs are path-connected $k$-vertex graphs. E.g., the cluster in Fig. \ref{cpcrF3}(a) belongs to $\C_3$. 
Then one has (\cite{Cam}, \cite{BGJ}) 
\beq 
\ln \tilde{Z}_\Lambda = \sum_{k=1}^\infty  \; {1\over k!} \sum _{X \in {\C}_k} \; q(X) \prod_{\gamma \in X} \zeta(\gamma)  \; , \label{logZ1}
\eeq
where the index $q(X)$ of the connected cluster $X$ is given by 
\beq
q(X) = \sum_{G_c\; {\rm on}\; X}  (-1)^{l(G_c)} \,. \label{logZ2}
\eeq
In (\ref{logZ2}) the sum is over the graph and all connected subgraphs $G_c$ on $X$, and 
$l(G_c)$ denotes the number of lines in $G_c$. Thus, e.g., the cluster in Fig. \ref{cpcrF3} has 
index $q(X) = -1 + 3\times 1= 2$.

Expectations (\ref{exp1}) are correspondingly given by 
\beq 
\vev{{\cal O}} =  
 \sum_{k=1}^\infty  \; {1\over k!} 
 \sum_{ X\in \,{\C}_k[{\cal O|}}
   \; q(X) \,\prod_{\gamma \in X} 
 \zeta(\gamma)  \; . \label{expexpan}
\eeq
In (\ref{expexpan}) ${\C}_k[{\cal O}]$ denotes the set of all connected $k$-clusters containing 
$(k-1)$ polymers with activities (\ref{polactiv1}) and 
one polymer $\gamma_{\ssc\cal O}$ 
having non-empty intersection with $O$ (the support of ${\cal O}$) 
with activity given by 
\beq 
\zeta(\gamma_{\ssc \cal O}) = \int \prod_{{\bf x}\in O\cup \gamma_{\ssc \cal O}} d\mu_{\bf x} 
\prod_{b_s \in \,\gamma_{\ssc \cal O}} dU_{b_s} \; \;{\cal O} \prod_{b_s, l,\alpha \, \in \,\gamma_{\ssc \cal O} } f^{l,\alpha}_{b_s} 
 \; .\label{polactivop} 
\eeq
(\ref{expexpan}) is easily obtained from (\ref{logZ1}) by writing 
\beq 
\vev{{\cal O}} = {Z_\Lambda[{\cal O}] \over Z_\Lambda} = 
{d\over d\lambda}\ln \tilde{Z}_\Lambda [(1+\lambda {\cal O})] 
\big\arrowvert_{\lambda=0} \,, \label{exp}
\eeq
where  $Z_\Lambda[{\cal O}]$ denotes the partition function with the insertion of the operator 
${\cal O}$.  

\subsection{Convergence} 
Having formulated our expansion as a polymer expansion for the logarithm of the partition function (\ref{logZ1}) or  for observable expectations (\ref{expexpan}) one may proceed to examine its  convergence by a straightforward application of known results  \cite{Cam}, \cite{BGJ} concerning the convergence of such expansions. 
As it is well-known, for translation invariant systems, the following convergence criterion holds 
\cite{F5}: 
polymer expansions (\ref{logZ1}) and  (\ref{expexpan}) converge absolutely and uniformly if 
\beq
Q\equiv  \sup_x \sum_{\gamma \ni x} \,|\zeta(\gamma)| \, e^{|\gamma|} \;  < \; 1  \,.  \label{convcrit1}
\eeq
In (\ref{convcrit1}) the sum is over all polymers $\gamma$ holding fixed a site $x\in \gamma$. 

From (\ref{prop3a})-(\ref{prop3b}) the propagator color-flavor matrix elements are bounded by 
\beq 
|C(\tau - \tau^\prime, \Theta(\bx))| \leq {2\over [1-e^{\D - \mu L}] }\,  e^{\D -\mu |\tau-\tau^\prime|} 
\, \label{propbound}
\eeq
with $|\tau-\tau^\prime|$  taken as the ``periodic" (shortest) distance between the two points in the timelike direction on the lattice torus.  From (\ref{Det}) one has 
\beq 
2^{-\nu L} e^{ \nu\mu L} \Big[ 1 -  e^{\D - \mu L}\Big]^{2\nu}
\leq   |{\rm Det} C^{-1}(\Theta(\bx))| \leq   2^{-\nu L} e^{ \nu\mu L} \Big[ 1 +  e^{\D - \mu L}\Big]^{2\nu} \,. \label{Detbound}
\eeq 
Consider a polymer $\gamma$ build out of a set of $|\gamma|$ spacelike bonds with bonds of multiplicity $m$ being counted $m$ times. (Note that in fact $m\geq 2$  because of the mod $N_c$ equal number of $U_b$ and $U_b^\dagger$'s integration constraint above.) 
Now, $U_{b_s}$, $U_{b_s}^\dagger$ are bounded by unity in color space. 
There are at most $\nu$ choices for connecting a $\bpsi$ or $\psi$ at a site $(\bx, \tau)$ on the boundary of one bond to  a $\psi$ or $\bpsi$ at a boundary site $(\bx, \tau^\prime)$ of another bond in the set via a propagator $C(\tau-\tau^\prime)$. There are $|\gamma|$ propagators in the diagram.  (Each propagator connects two bonds but each bond has two boundary sites; also, note that only sites separated by an odd number of bonds in the timelike direction can be so connected.)
From (\ref{polactiv1})-(\ref{polactiv2}), (\ref{z}) and (\ref{propbound}), (\ref{Detbound}) then one has 
\beq 
|\zeta(\gamma)| \leq (\nu)^{2|\gamma|} \left[{1 +  e^{\D - \mu L} \over 1 -  e^{\D - \mu L}}\right]^{2\nu |\bx|}   \left[{2 \over 1 -  e^{\D - \mu L}}\right]^{ |\gamma|} 
\prod_{{\bf x}\in \gamma} \prod_{(\tau_i \tau_j)}   e^{-\D \mu |\tau_i-\tau_j|} \, . \label{activbound}
\eeq
To next sum over all such polymers made of $|\gamma|$ spacelike bonds and attached to a fixed site we  sum over all possible timelike separations between the bonds and over all possible configurations of  the bonds. Now 
\beq
\sum_{\tau=1}^{L/2} e^{\D - \mu|\tau|} <  e^{\D - \mu} [1 + \sum_{\tau=1}^{L-1} 
e^{\D - \mu|\tau|}]=  e^{\D - \mu}\left[{ 1- e^{\D - \mu L} \over 1- e^{\D - \mu}}\right] \,,
\label{propsum} \nonumber 
\eeq
whereas the number of possible spacelike bond configurations is easily seen to be bounded by 
$(2d)^{2|\gamma|}$. 
Combining with (\ref{activbound}) and noting that $|\bx|\leq |\gamma|/2$ one finally has 
\beq 
Q < \sum_{|\gamma|=1}^\infty K^{|\gamma|} (e^\mu -1)^{-|\gamma|} e^{|\gamma|} = 
{eK\over e^\mu -1 - eK}\, ,  \label{Qsum} 
\eeq 
where 
\beq 
K\equiv 2\nu^2 (2d)^2 \left[ { 1+ e^{\D - \mu L} \over 1 - e^{\D-\mu L} } \right]^\nu 
\,. \label{Kdef} 
\eeq
The convergence condition $Q<1$ then gives $2eK < (e^\mu - 1)$. 
We have thus arrived at the following result. {\it For any spatial dimension $d\geq 1$ the polymer expansions (\ref{logZ1}) and (\ref{expexpan}) converge absolutely and uniformly in $|\Lambda_s|$ 
if }
\beq 
4e\nu^2 (2d)^2 < [\tanh(\mu /2a_{\tau}T)]^\nu\,( e^{\D \mu} - 1)
\,. \label{convcrit2}
\eeq
In particular, one has:  
\beq
\ln[1 + 4e\nu^2 (2d)^2] < \mu  \qquad \mbox{for} \qquad T=0 \,. \label{convcrit3}
\eeq
A sufficient condition on $\mu$ satifying (\ref{convcrit2}) for any temperature ($2 \leq L < \infty$) is in fact given by 
\beq 
\ln\left[1 + 4e\nu^2 (2d)^2 \left( \coth( \ln[1+ 4e\nu^2 (2d)^2])\right)^\nu \right]< \mu  \,. \label{convcrit4}
\eeq
\begin{figure}[ht]
\begin{center}
\includegraphics[width=0.3\textwidth]{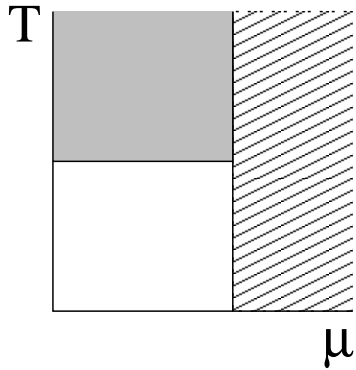}
\end{center}
\caption{ Striped region of convergence (chiral symmetry)  given by (\ref{convcrit4}) in text. 
The complete chirally symmetric region is expected to include the shaded area connecting to the region of low $\mu$ at high $T$.  \label{cpcrF4}} 
\end{figure}

It should be noted that (\ref{convcrit2}) is a substantial overestimate since it was arrived at by ignoring, for the most part, the spacelike bond multiplicity constraints as well as all factors of inverse powers of $N_c$ resulting from the spacelike gauge field integrations and the restrictions to odd timelike separations. The bound improvements from taking these into account, however, do not change the qualitative picture 
represented by Fig. \ref{cpcrF4}. More importantly, one may expect that the high $\mu$ convergence region connects to the high $T$ convergence region \cite{TY} (through the shaded area in  Fig. \ref{cpcrF4}). 
To establish this one first needs to reintroduce arbitrary $(a_t/a_s)$  in the spacelike action. This ratio was conveniently set equal to one above but it provides crucial convergence factors at large $T$ and small $\mu$. After this trivial step, one needs to bound the gauge field $\Theta$ dependence (at small $\mu$) less crudely than in (\ref{propbound}) and (\ref{Detbound}).

\section{Chiral symmetry} 
\setcounter{equation}{0}
\setcounter{Roman}{0}
An immediate consequence of the convergence of the cluster expansion (\ref{expexpan}) is the existence of chiral symmetry within the region of its convergence. 

Indeed, the expectation of any local chirally non-invariant fermion operator ${\cal O}(x)$, such as, e.g., $\bpsi(x)\psi(x)$, vanishes identically  term by term in the expansion (\ref{expexpan}) by the chiral invariance of the measure $d\mu_\bx$. Correlation functions $\vev{{\cal O}(x) {\cal O}(y)}$ then receive vanishing contributions from any polymers consisting of two disjoint clusters connected to sites $x$ and $y$, respectively. Nonvanishing contributions  arise only from polymers intersecting  both sites $x$ and $y$. Now, the total number $n_T$ of timelike bonds in propagators in any minimal such polymer is easily seen to be bounded from below by the minimal distance between the two sites. 
Also, there are  $|\gamma|$ propagators in a polymer $\gamma$, and $|\gamma|\leq n_T$ since each 
propagator is at least one timelike bond long. 
From (\ref{activbound}) then the activity of such a minimal polymer is bounded from above by 
\bea 
& &  \left[ (\nu)^2 \left[{1 +  e^{\D - \mu L} \over 1 -  e^{\D - \mu L}}\right]^\nu  \left({2 \over 1 -  e^{\D - \mu L}}\right)  \right]^{\D n_T}   e^{\D -\mu n_T}  \nonumber \\
 & = & \left[ {K\over (2d)^2}\left( {1 \over 1 -  e^{\D - \mu L}}\right)  \right]^{\D n_T}   e^{\D -\mu n_T}  
  <   \left[ \left({1 - e^{\D -\mu} \over 1 - e^{\D -\mu L}}\right) {1 \over 2e (2d)^2} \right]^{\D n_T} \leq  \left[ {1 \over 2e (2d)^2} \right]^{\D n_T}   \nonumber 
 \eea  
 where we used the convergence condition $2eK < (e^\mu - 1)$ from above. From this it follows that within the convergence radius of the expansion 2-point correlations are absolutely and uniformly bounded at any temperature from above by 
\beq 
\left|\vev{{\cal O}(x) {\cal O}(y)}\right|  < {\rm Const.} 
\left[ {1 \over 2e (2d)^2} \right]^{\D |x-y|}  \, , \label{corrbound}
\eeq
where $|x-y|$ is the minimum number of bonds connecting the two sites, i.e., 
\beq
 \lim_{|x-y|\to \infty}\lim_{|\Lambda_s| \to \infty} \vev{{\cal O}(x) {\cal O}(y)}  \to 0 \,.  
 \label{corrcluster}
 \eeq
In the same manner, all higher correlation  functions of any chirally non-invariant fermion operators  
(\ref{exp2}) exhibit exponential clustering for large separations, i.e., there is no spontaneous breaking of the global chiral symmetry.  

We have thus shown the existence of a chirally symmetric phase at sufficiently large chemical potential $\mu$ (and any temperature $T$). 
Chiral symmetry in $SU(N_c)$ gauge theories at vanishing $\mu$ and $T$ is of course spontaneously broken. This breaking, studied in a vast literature over several decades, was 
established analytically in  lattice gauge theories at strong coupling by hopping expansion, large $N_c$, and other techniques, long ago \cite{BBEG}. Our result then implies the existence of a chiral symmetry restoring phase transition at some critical $\mu$, as indeed seen in the studies cited in section 1. 

An interesting issue in this connection concerns the dependence on the number of fermion flavors. 
Recently, it was found that, contrary to previous belief, even in the strong coupling limit at vanishing $T$ and $\mu$, 
chiral symmetry is restored if the number of flavors becomes large enough (for fixed $N_c$). This surprising result was established by numerical simulations for $SU(3)$ gauge group \cite{dFKU}, and, more generally, for $SU(N_c)$ by resummations of fermion hopping expansions \cite{T}.  
Now, in our estimates (\ref{convcrit2}),   (\ref{convcrit4}) for $\mu$ such that chiral symmetry is present the flavor dependence is contained in $\nu$. These estimates increase with $\nu$. This is simply because, in proving absolute convergence for large $\mu$, they were obtained by bounding the absolute value of each term in the expansion from above. 
This completely obscures the mechanism of chiral restoration at large $N_f$ which relies on 
cancelations between classes of diagrams  as $N_f$ is varied (at fixed $N_c$): subdominant graphs at small $N_f$ become dominant at large $N_f$ with such signs as to destroy the condensate formed by the graphs dominant at small $N_f$ \cite{T}. Having established absolute convergence at large $\mu$ as above, it may be possible to group and resum terms in our expansion so as to exhibit such cancellations along the lines in \cite{T}. This 
would, however, require rather more elaborate estimates (including the inverse powers in $N_c$ 
mentioned in the last paragraph of the previous section) than the straightforward graph by graph bounds above, and will not be pursued here. If carried out, however, one would expect a result qualitatively as in 
Fig. \ref{cpcrF4} with $N_f$, instead of $T$, labeling the vertical axis.

\section{Conclusion}
\setcounter{equation}{0}
\setcounter{Roman}{0}

We have shown that for sufficiently large quark chemical potential strongly coupled $SU(N_c)$ lattice gauge theories with $N_f$ massless staggered fermions and at any temperature are in a chirally symmetric phase.   
This agrees with and generalizes previous results obtained by numerical or mean field techniques in particular cases as reviewed in section 1. 

There are several directions in which this work may be further pursued. 
The lower bounds on the magnitude of the critical chemical potential given in section 2.3 are very non-optimal and can be improved as outlined. As also mentioned there, it should be possible 
to extend the region of convergence in the $\mu-T$ plane to high $T$ and small $\mu$, thus connecting with the high $T$ proof in \cite{TY}. This would require doing a 
better job of bounding timelike gauge field integrations at small $\mu$, and 
would appear quite feasible, especially in particular cases such as $N_c=2,3$. 

Such improvements may be viewed within the general framework of employing   
the expansion as a computation tool. In this paper we were mostly concerned with  establishing the convergence of the expansion rather than its computational use. 
Explicit computation of graphs involves integration over the gauge fields on spacelike bonds in the hopping expansion and integration over the timelike gauge field dependence in the propagators. 
The former, familiar from ordinary hopping and strong coupling expansions on the lattice, can be carried out systematically to quite high order. The latter will in general require numerical evaluation. 
Truncating the series to leading approximation will give, as already mentioned, results quite close 
to those in \cite{NFH} - \cite{KMOO}. The expansion then provides systematic corrections to such leading approximations. As in the case of ordinary hopping expansions, one might hope that resummation of (infinite) classes of diagrams may also allow extension beyond the radius 
of convergence of the original series, and/or improved mean field or other approximations in a wider regime. The effects of increasing number of flavors, discussed at the end of the previous section, may also be exhibited by such resummations.

A  physically interesting and obvious question is whether the present treatment can be extended 
from the strong coupling limit to the finite gauge coupling region. 
For small enough inverse gauge coupling $\beta$ this is perfectly feasible by extending the large $\mu$ expansion above to incorporate the well-known and well-studied small $\beta$ cluster expansion (e.g., \cite{Mu}).  
The expansion of the exponential of the plaquette gauge action contributes correction terms proportional to powers of $\beta$ that ``embellish" 
the graphs in our expansion by the familiar plaquette corrections to a  
fermion hopping expansion - the spacelike bond integrations being as before and, in the present finite $\mu$ application,  the remaining timelike bond integrations in the measure provided by the timelike propagators.  The straightforward inclusion of  the 
plaquette corrections \cite{Mu} to our previous graph counting and bounds  gives then 
convergence for sufficiently large $\mu$ and small $\beta$. 
For general, and in particular large $\beta$ values, however, where no convergent expansion 
techniques in the plaquette interactions are available, extension of the present results appears problematical. This is certainly true in the case of odd number of colors $N_c$. 
For even $N_c$, however, and in particular for $N_c=2$, the real positive nature of the fermionic determinants (cf. (\ref{Detsu2})) may allow one to ``repackage" the above expansion making use of this positivity to extend into the finite coupling region at large $\mu$. We hope to explore this elsewhere. 

\vspace{0.5cm} 

The author would like to thank P. de Forcrand for correspondence. 

\setcounter{equation}{0}
\appendix
\renewcommand{\theequation}{\mbox{\Alph{section}.\arabic{equation}}}

\section{Appendix: Derivation of some formulas in the main text}
In this Appendix we derive the explicit expressions (\ref{prop3a}), (\ref{prop3b}), (\ref{proprel}) and (\ref{Det}). 
In the following we set $\gamma_0=1$ in (\ref{prop2})  for staggered fermions. We always take even $L$. 

Given a meromorphic function $g(z)$ which is analytic on $|z|=1$, summation over the Brillouin zone (\ref{BZ}) can be performed by using the easily verified representation 
\beq
{1\over L} \sum_{k_m\in {\sf BZ}} g(e^{ik_m}) = - {1\over 2\pi i}\, (\oint_{|z|=1+\epsilon} - 
\oint_{|z|=1-\epsilon} ) \;{dz\over z} {g(z) \over [z^L+ 1]} \label{contrep}
\eeq
(both contours taken counterclockwise). Now if the integrand tends to zero as $|z|^{-2}$ or faster as $|z|\to \infty$ one may deform the outer contour to infinite radius taking into account any poles of $g(z)/z$ encountered in this deformation. The integral over the inner contour may be performed by residues of its enclosed poles.

We use (\ref{contrep}) to evaluate the propagator (\ref{prop1}) - (\ref{prop2}), which, written out separately for $\tau>0$ and $\tau < 0$, has the form
\beq 
 C(\tau, \Theta(\bx))_{ai,bj} = \mp i\delta_{ab}\delta_{ij} {1\over L} \sum_{k_m\in {\sf BZ}}
 { e^{\D i k_m |\tau| } \over 
\sin[k \pm \theta_a(\bx)/L \mp i\mu]} \qquad \mbox{for} \quad \tau= \pm |\tau| \, .  \label{Aprop1}
\eeq

For $\tau >0$ and $\mu >0$ one then obtains 
\bea
 C(\tau, \Theta(\bx))_{ai,bj}&  = & \delta_{ab}\delta_{ij}{i\over \pi} e^{ -i\theta_a(\bx)/L}e^{-\mu}   \nonumber \\
& & \  \cdot \; (\oint_{|z|=1+\epsilon} - \oint_{|z|=1-\epsilon} ) \;dz \; {\D z^\tau \over 
\D [z^2 - e^{ -i2\theta_a(\bx)/L}e^{-2\mu} ]}{1\over [1 +z^L]}\,. \label{Aprop2}
 \eea
Now, since $|\tau|<L$, the integrand is sufficiently convergent for the outer contour to be deformed to infinite radius where it gives zero contribution. No poles are encountered in this deformation,  the integrand having poles at  
$z_{\pm} = \pm e^{ -i\theta_a(\bx)/L}e^{-\mu}$ inside the inner contour. Evaluation of the inner integral by residues gives then (\ref{prop3a}). 

For $\tau <0$ and $\mu >0$ one has from (\ref{Aprop1}) 
\bea
 C(\tau, \Theta(\bx))_{ai,bj}&  = & - \delta_{ab}\delta_{ij}{i\over \pi} e^{ i\theta_a(\bx)/L}e^{\mu}   \nonumber \\
& & \  \cdot \; (\oint_{|z|=1+\epsilon} - \oint_{|z|=1-\epsilon} ) \;dz \; {\D z^{|\tau|} \over 
\D [z^2 - e^{ i2\theta_a(\bx)/L}e^{2\mu} ]}{1\over [1 +z^L]}\,. \label{Aprop3}
 \eea
The inner contour integral now encloses no singularities and gives zero contribution.  The outer contour can again be deformed to infinite radius, now, however, encountering the integrand singularities at $z_{\pm} = \pm e^{ i\theta_a(\bx)/L}e^{\mu}$. Evaluating their contribution results into (\ref{prop3b}). The two cases are reversed for $\mu <0$ (antiquark chemical potential). 

An alternative method of summation over the Brillouin zone is by use of the 
Fourier series representation of $\sum_{m= - \infty}^\infty \delta(x- 2m\pi)$: 
\bea
{1\over L} \sum_{k_m\in {\sf BZ}} e^{ik_m\tau} g(k_m) & = & 
{1\over L} \sum_{m=-(L/2) + 1}^{L/2}  \int_{-\pi}^\pi dk \, \delta(k- k_m) \, e^{ik\tau} g(k) \nonumber \\
& = &  \sum_{m=-\infty}^\infty  \int_{-\pi}^\pi dk \,\delta(kL- (2m-1)\pi)\,  e^{ik\tau} g(k)  \nonumber \\
& = & \sum_{n=-\infty}^\infty  \int_{-\pi}^\pi {dk\over 2\pi}\,  e^{i n (kL + \pi)} 
 e^{ik\tau} g(k)   \nonumber \\
& = & \sum_{n=-\infty}^\infty  (-1)^n \, \left[ \int_{-\pi}^\pi {dk\over 2\pi}\, 
e^{ik(\tau + nL)} g(k) \right] \,.   \label{AdelFSrep} 
\eea
Applying (\ref{AdelFSrep}) to (\ref{prop1}) one obtains 
\beq 
 C(\tau, \Theta(\bx)) = \sum_{n=-\infty}^\infty  (-1)^n \,  \tilde{C}(\tau + nL, \Theta(\bx)) 
  \label{Aprop4}
 \eeq
with $\tilde{C}(\tau, \Theta(\bx))$ given by (\ref{tprop}), i..e, by the propagator (\ref{prop1}) in the $L\to \infty$ limit of the {\sf BZ}. 
For $\tau >0$ and $\mu>0$, letting $z=e^{ik}$, (\ref{tprop}) may be represented in the form: 
\beq
\tilde{C}(\tau, \Theta(\bx))_{ai,bj} = - \delta_{ab}\delta_{ij} {i\over \pi} \,
e^{ -i\theta_a(\bx)/L}e^{-\mu} \,  \oint_{|z|=1} \;dz \; {\D z^\tau \over 
\D [z^2 - e^{ -i2\theta_a(\bx)/L}e^{-2\mu} ]} \,.  \label{Atprop1}
\eeq
Evaluating the contour integral gives:
\beq 
\tilde{C}(\tau, \Theta(\bx))_{ai,bj} =  \delta_{ab}\delta_{ij} [1 - (-1)^{\D \tau}] \;
e^{\D -i\theta_a(\bx)\tau/L} e^{\D -\mu \tau}\, ,  \qquad    \tau > 0, \quad  \mu>0 \, . \label{Atprop2}
\eeq
For $\tau= - |\tau| < 0$ and $\mu>0$ one obtains 
\bea 
\tilde{C}(\tau, \Theta(\bx))_{ai,bj}& = &   \delta_{ab}\delta_{ij} {i\over \pi} \,
e^{ i\theta_a(\bx)/L}e^{\mu} \,  \oint_{|z|=1} \;dz \; {\D z^{|\tau|} \over 
\D [z^2 - e^{ i2\theta_a(\bx)/L}e^{2\mu} ]}   \label{Atprop3} \\
& = & 0  \, ,   \qquad    \tau < 0, \quad  \mu>0  \, .  \label{Atprop4}
\eea
This is as expected since, at fixed $\bx$,  (\ref{tprop}) represents the propagator of a 1-dimensional static (no kinetic energy term) fermion in the background of a constant potential. 
For $\mu <0$, (\ref{Atprop2}) and (\ref{Atprop4}) are reversed (antifermion propagation).  

Since $L > |\tau|$, (\ref{Atprop4}) reduces (\ref{Aprop4}) to 
(\ref{proprel}), i.e., 
\beq 
C( \tau, \Theta(\bx)) =\sum_{n=0}^\infty  (-1)^n \tilde{C}( \tau +nL, \Theta(\bx))  \,. 
\label{Aproprel} 
\eeq 
(The $n=0$ term is actually absent for $\tau <0$). 
Inserting the explicit form (\ref{Atprop2}) in (\ref{Aproprel}) then and carrying out the sum 
for $\tau = \pm |\tau|$ results in  (\ref{prop3a}) and (\ref{prop3b}), respectively.

Integrating out the fermions in (\ref{dmux}), i.e., along a one-dimensional timelike chain at fixed $\bx$ with  action (\ref{dmuxact2}) gives 
\beq
 {\rm Det} C^{-1}(\Theta(\bx)) =  
 \left[\prod_{a=1}^{N_c} \prod_{m=-L/2+1}^{L/2}i \sin\left({(2m-1)\pi \over L} + \theta_a/L -i\mu\right) \right]^{N_f} \, . \label{ADet1}
\eeq
Using the identity 
\beq 
\prod_{n=0}^{L-1} \sin\left( {2n\pi\over L} + z\right) = {(-1)^{L/2}\over 2^{L-1}} ( 1- \cos (Lz))
\label{trigid} 
\eeq
(\ref{ADet1}) becomes 
\bea  
{\rm Det} C^{-1}(\Theta(\bx)) & = & i^{L\nu} (-1)^{L\nu/2}2^{-L\nu} 
 \left[\prod_{a=1}^{N_c} \left[ 2 + e^{i\theta_a} e^{\mu L} + e^{-i\theta_a} e^{-\mu L}
\right]\right]^{N_f} \nonumber \\
& = & 2^{-\nu L} e^{\nu \mu L}
 \left[\prod_{a=1}^{N_c} \left[ 1 +  e^{-i\theta_a} e^{-\mu L}\right]^2\right]^{N_f} 
 \label{ADet2} \, . 
 \eea
 (\ref{Polgauge3}) was used in obtaining (\ref{ADet2}) which is (\ref{Det}) of the main text.


\begin{thebibliography}{99}
\bibitem{FH} K. Fukushima and T. Hatsuda, Rep. Prog. Phys. 74, 014001 (2011) [arXiv:1005.4814 [hep-lat]]
\bibitem{dF} P. de Forcrand  PoS LAT2009, 010 (2009)    [arXiv:1005.0539]. 
\bibitem{DMW} E. Daggoto, A. Moreo and U. Wolff, Phys. Lett. B 186, 395 (1987). 
\bibitem{KM} F. Karsch and K. H. M\"{u}tter, Nucl. Phys. B {\bf 313}, 541 (1989). 
\bibitem{KIM} J. U. Klaetke and K. H. M\"{u}tter, Nucl. Phys. B {\bf 342}, 764 (1990). 
\bibitem{CJ} S. Chandrasekharan and F-J. Jiang, Phys. Rev. D {\bf 74}, 014506 (2006) [arXiv:hep-lat/0602031].
\bibitem{AC} D. H. Adams and S. Chandrasekharan, Nucl. Phys. B {\bf 662}, 220 (2003) 
[arXiv:hep-lat/0303003]. 
\bibitem{FdF} M. Fromm and P. de Forcrand, PoS LAT2008, 191 (2008) [arXiv:0811.1931]. 
\bibitem{NFH}Y. Nishida, K. Fukushima and T. Hatsuda, Phys. Rept. {\bf 398}, 281 (2004) [arXiv:hep-ph/0306066].
\bibitem{N} Y. Nishida, Phys. Rev. D {\bf 69}, 094501 (2004) [arXiv:hep-ph/031231]. 
\bibitem{KMOO} N. Kawamoto, K. Miura, A. Ohnishi and T. Ohmura, Phys. Rev. D {\bf 75}, 014502 (2007) [arXiv: hep-lat/0512023]. 
\bibitem{TY} E. T. Tomboulis and L. G. Yaffe, Comm. Math. Phys. 100, 313 (1985); Phys. Rev. Lett. {\bf 52}, 2115 (1984). 
\bibitem{F0} Reflection positivity in timelike planes, which was invoked in the convergence proof of the type of cluster expansion used in \cite{TY},  no longer holds  in the presence of nonvanishing chemical potential. 
\bibitem{Cam} C. Cammarota, Comm. Math. Phys. 85, 517 (1982). 
\bibitem{BGJ} D. C. Brydges, in {\it Critical Phenomena, Random Systems and Gauge Theories},  
Les Houches XLIII, K. Osterwalder  and R. Stora (eds.), Elsevier (1986). 
\bibitem{F01} Whatever (non)convergence properties this $1/d$ expansion may have at physical values of $d$, its truncation to leading terms mimics the corresponding truncation in the expansion considered in this paper.  
\bibitem{F1} More generally, the global symmetry is $G\times G$ where 
$G$ is the subgroup of $U(N_f)$ which commutes with all the matrices $\{ \gamma[b]\}$. 
For staggered fermions $G=U(N_f)$. 
\bibitem{Fim} On a more technical note, we do not employ any sort of ``improved" staggered fermion actions. Indeed, such ``improvements" are only designed to assist approach to the continuum limit. They are not suited for our application at strong coupling, finite lattice spacing, where they are generally known to lead to unphysical effects such as spurious phases and other artifacts due to explicit violation of reflection positivity and other undesirable features. 
\bibitem{F2a} Thus, with parametrization $U=\exp(i \mbox{\boldmath $\omega$} \cdot {\bf h})$, where 
${\bf h} = (h_1, \ldots, h_{N-1})$ are the $SU(N)$ Cartan subalgebra generators,  one may set $\theta_i = \mbox{\boldmath$\omega$} \cdot{\bf  m}_i $, where ${\bf m}_i$ are the fundamental representation root vectors. 
\bibitem{F2b} One may note in this connection that in 
the formal continuum theory topological obstructions may arise in attaining the Polyakov gauge: 
C. Ford, T. Tok, A. Wipf, Phys. Lett. B 456, 155 (1999); E. Langmann, M. Salmhofer, A. Kovner, Mod. Phys. Lett. A9, 2913 (1994). 
The resulting singularities define the location of singular strings, monopoles, etc. Such singularities cannot, of course, appear in the lattice-regularized theory. 
\bibitem{F3} Indeed, each connected component of a given set $B$ may give rise to more than one distinct connected diagram from distinct Wick theorem contractions embodied in the integrations over 
$d\mu_{\bf x}$. Different connected components of a given set $B$ give of course rise to disjoint connected diagrams. 
\bibitem{F4} Here the standard graph theory definition of ``graph" is used: a set of vertices connected by lines (edges), with any two vertices connected by at most one line. 
\bibitem{F5} The well-known basic idea of the proof, introduced in \cite{Cam}, is to replace the sum over 
$\C_k$ in (\ref{logZ1}), (\ref{expexpan}) by a sum over trees and use Cayley's theorem on the number of trees of given coordination numbers, then sum over all coordination numbers. The convergence criterion in the concise form (\ref{convcrit1}) is obtained in \cite{BGJ}. 
\bibitem{BBEG} J-M Blairon, R. Brout, F. Englert and J. Greensite, 
Nucl. Phys. {\bf B 180}[FS2], 439 (1981); 
O. Martin and B. Siu, Phys. Lett. B 131B, 419 (1983); 
N. Kawamoto and J. Smit, 
 Nucl. Phys. {\bf B 192}, 100 (1981); H. Kluberg-Stern, A. Morel, O. Napoly and B. Peterson, 
 Nucl. Phys. {\bf B 190} [FS3], 504 (1981). 
 H. Kluberg-Stern, A. Morel and B. Peterson, 
 Nucl. Phys. {\bf B 215} 527 (1983). 
 For $U(N)$ gauge theories, where, however, a baryon chemical potential cannot exist, a rigorous proof of chiral symmetry breaking at strong coupling and zero temperature was given in M. Salmhofer and E. Seiler, Comm. Math. Phys., 139, 395 (1991). 
 
\bibitem{dFKU} Ph. de Forcrand, S. Kim and W. Unger, JHEP 02(2013)051 [arXiv:1208.2148]. 
\bibitem {T} E. T. Tomboulis, Phys. Rev. D {\bf 87}, 034513 (2013) [arXiv:1211.4842]. 
 \bibitem{Mu} G. M\"{u}nster, Nucl. Phys. {\bf B 180} [FS2], 23 (1981). 

\end{thebibliography}
\end{document}